\documentclass[preprint,aps,eqsecnum,amsmath,nofootinbib]{revtex4}
\usepackage{graphicx,epsfig}
\usepackage{amsmath}
\usepackage{indentfirst}
\usepackage{hyperref}
\usepackage{bm}
\usepackage{ifpdf}
\newcommand{\nn}{\nonumber}

\def\nn{\noindent}

\def\to{\rightarrow}

\def\be{\begin{equation}}
\def\ee{\end{equation}}
\def\bqa{\begin{eqnarray}}
\def\eqa{\end{eqnarray}}
\begin{document}\baselineskip 20pt
\rightline{\vbox{\halign{&#\hfil\cr &GUCAS-CPS-07-007 \cr &MIT-CTP
3911 \cr &hep-ph/0710.3339\cr \cr\cr\cr}}}

\title{Investigation of the bottomonium ground state $\eta_b$ via its inclusive charm decays\\[8mm]}

\author{Gang Hao$^1$\footnote{Email:
hao$_-$gang@mails.gucas.ac.cn}, Cong-Feng
Qiao$^{1,2}$\footnote{Email: qiaocf@gucas.ac.cn}, Peng
Sun$^1$\footnote{Email: sunpeng05@mails.gucas.ac.cn}}
\address{\vspace{5mm}
{\rm (1)} Department of Physics, Graduate University of Chinese
Academy of
Sciences,\\
YuQuan Road 19A, Beijing 100049, China \\
{\rm (2)} Center for Theoretical Physics, LNS and Department of Physics\\
Massachusetts Institute of Technology, Cambridge, Massachusetts
02139, U.S.A. \vspace{18mm}}

\begin{abstract}\vspace{5mm}

Based on the non-relativistic QCD factorization formalism, we
calculate the bottomonium ground state, $\eta_b$, inclusive charm
decays at the leading order in the strong coupling constant
$\alpha_s$ and quarkonium internal relative velocity $v$. The
inclusive charm pair production in $\eta_b$ decay is mainly realized
through $\eta_b \rightarrow c\ \bar{c}\ g$ process, where the charm
and anti-charm quarks then dominantly hadronize into charmed
hadrons. The momentum distributions of the final states are
presented. In this work, we also calculate the $J/\psi$ inclusive
production rate in the $\eta_b$ decays, where the color-octet
contribution is found to be very important. We expect this study may
shed some light on finding $\eta_b$ or knowing more about its
nature.
\\[1mm]

\noindent PACS number(s): 12.38.Bx, 12.39.Hg, 13.20.Gd
\end{abstract}
\maketitle


\section{Introduction}
In high energy research, heavy quarkonium physics is one of the most
interesting fields and it plays an important role in understanding
the hadron configuration and the microcosmos. Theoretically, due to
the non-relativistic nature of heavy quarknoia, it is convenient to
research their properties in the framework of non-relativistic
potential model and non-relativistic QCD \cite{NRQCD}, or other
theories which work well in the non-relativistic limit.
Experimentally, heavy quarkonia have relatively high production rate
in both electronic and hadronic collisions and the vector members
can be easily seen through their bi-lepton decays. Recently, some
new resonances have been observed in the charmonium energy region
\cite{swanson}, which enrich the heavy quarkonium spectroscopy and
make heavy-quarkonium physics more interesting.

After the spin-triplet bottomonium $\Upsilon$ was discovered three
decades ago, its pseudoscalar partner $\eta_b$ had been looked for
in various experiments. Unfortunately, there is still no conclusive
evidence that this elusive particle has been found. As a solid
prediction from the quark model, the existence of $\eta_b$ is
indubitable, but its mass and decay channels remain undetermined
experimentally at the present time. To search $\eta_b$, several
experiments have been conducted both in $e^+e^-$ collisions at the
CLEO and the LEP, and hadronic collisions at the Fermilab Tevtron.
The advantage of $e^+e^-$ collisions is its clear background, which
is impaired by the fact that production rates for spin-singlet
states are generally small. Based on the 2.4 ${\rm fb}^{-1}$ data
taken at the $\Upsilon(2S)$ and $\Upsilon(3S)$ resonances, CLEO has
searched distinctive single photons from hindered $M1$ transitions
of $\Upsilon(2S)$ and $\Upsilon(3S)$ to $\eta_b\gamma$, and also
from the cascade decay $\Upsilon(3S)\to h_b\pi^0,\:h_b\pi^+\pi^-$
followed by $E1$ transition $h_b\to \eta_b\gamma$, but no signals
have been found \cite{cleo1}. In the experiments at LEP II, the
$\eta_b(1S)$ is expected to be produced in two-photon process. The
ALPHA Collaborations analyzed the $\gamma\gamma$ interaction data,
but found no evident signal in the four- and six-charged-particle
final states \cite{ALPHA}. The results of the L3 Collaboration and
the DELPHI Collaboration were also negative and the upper limits of
the products $\Gamma_{\gamma\gamma}\times Br(\eta_b)$ were set
\cite{L3,delphi1,delphi2}. In comparison to $e^+\ e^-$ experiments,
the hadronic collision at the Fermilab Tevatron gives large $\eta_b$
production rate, but due to the complicated hadronic interaction
background,  the search for $\eta_b$ there is also hard. Using the
full 1992-96(Run I) data, the CDF Collaboration searched the
$\eta_b$ through its exclusive decay to double $J/\psi$ and found
some oblique evidences, but far from conclusive \cite{cdf}. Right
now, further efforts is being pursued in the Run II data there.

Considering the situation of $\eta_b$ experiments, theoretical
research on its properties is still necessary, such as its mass, the
production cross-sections at different colliders, and its various
decay channels. Among all properties, the $\eta_b$ mass is believed
to be predicted by potential model, effective theory, and Lattice
calculation without much ambiguity, which is very important for
experimental observation. Recent theoretical work fixes the
$\Upsilon-\eta_b$ mass splitting in the range of $40-60$ MeV
\cite{mass1, mass2, mass3, mass4}. In Ref. \cite{product1}, Braaten,
Fleming and Leibovich calculated the $\eta_b$ production cross
section at the Fermilab Tevatron in the framework of NRQCD and
evaluated the branching ratio of the decay $\eta_b\rightarrow J/\psi
J/\psi$. They suggested that $\eta_b$ should be observable through
this decay because of its large branching ratio of $7\times
10^{-4\pm1}$. In Ref. \cite{product2}, Maltoni and Polosa also
evaluated the observation potential for $\eta_b$ at the Tevatron,
but they found and suggested that the decay $\eta_b\rightarrow D^*
D^{(*)}$ might be the most optimistic channel to observe $\eta_b$
signal. In Ref. \cite{jia}, the relativistic correction to the
$\eta_b\rightarrow J/\psi J/\psi$  decay process was calculated and
a much smaller branching ratio in comparison to Ref. \cite{product1}
was obtained. The author also discussed the $\eta_b\rightarrow D^*
D^{(*)}$ process and got a smaller rate than what in Ref.
\cite{product2}. However, Santorelli finds that the final state
interaction may enhance $\eta_b \rightarrow J/\psi J/\psi$ decay
width by about two orders of magnitude \cite{San}. Instead of
$\eta_b$ decays into hadronic final states, Hao {\it et al.}
calculated the branching ratio of the $\eta_b$ radiative decay
process, i.e. $\eta_b\rightarrow\gamma J/\psi$ \cite{qiao}. They
claimed that this channel is also a hopeful one in the $\eta_b$
hunting.

Comparing to the exclusive process, the inclusive process has a
large branching ratio, nevertheless normally also has large
uncertainties in pinning down the parent particle. Fortunately, the
final state distributions of experimental observables are helpful to
the inclusive process for the aim. For instance, recently, the
inclusive charm production in the $\chi_b$ and $\Upsilon$ decays
have been studied in \cite{CPX} and \cite{CPW} respectively.

At the leading order in $\alpha_s$ and non-relativistic expansion,
the $\eta_b$ inclusive open charm decay happens through $b\
\bar{b}\rightarrow g\ g^*$ followed by $g^*\rightarrow c\ \bar{c}$,
where the initial $b\bar{b}$ is configured in color-singlet. Based
on the result of $b\ \bar{b}\rightarrow g\ c\ \overline{c}$, we can
roughly estimate the branching fraction of $b\ \bar{b}\rightarrow X
+$ {\rm charmed hadrons}. We calculate this process in the framework
of NRQCD factorization formalism, in which the un-calculable
nonperturbative effects are represented by the matrix elements of
NRQCD operators. According to NRQCD, the color-octet configurations
appear as higher order Fock states in  $\eta_b$ decays, which are
suppressed by orders of the small magnitude in relative velocity
$v^2$. Hence, for a leading order calculation, we can safely treat
the $b\ \bar{b}$ pair inside the $\eta_b$ to be in color-singlet.
However, for the $J/\psi$ production in $\eta_b$ decays, although
the higher order Fock state, the $(c\bar{c})(^3S_1^{[8]})$, is
suppressed by $v^4$ relative to the leading color-singlet
configuration, it may be compensated by a factor of $\alpha_s$ in
regarding to color-singlet process.

This paper is organized as follows: the inclusive charm and charmed
hadron production in $\eta_b$ decay is evaluated at leading order in
Section \ref{section2}.  In Section \ref{section3}, calculation of
the process $\eta_b \rightarrow J/\psi+X$ will be presented. The
last section is remained for summary.

\section{the charm quark production in $\eta_b$ decays}\label{section2}

In this section we calculate the charm quark production in the
inclusive $\eta_b$ decays. At the leading order in $v$, according to
the NRQCD factorization formulism, its open charm decay width takes
the form
\bqa\ \Gamma [\eta_b\to c\ \bar{c} + X] = C_1^c\ \frac{\langle
\mathcal{O}_1(^1S_0)\rangle_{\eta_b}}{m_b^2}\ ,\eqa
where $\langle \mathcal{O}^{\eta_b}_1(^1S_0)\rangle$ is a NRQCD
matrix element, which represents the long-distance effect and gives
the probability for finding the heavy quark and antiquark in
specific configuration within the meson, and can be evaluated by
nonperturbative method such as lattice simulation. The dimensionless
short-distance coefficient $C_1^c$ can be calculated in pQCD. The
dominant source of $C_1^c$ comes from the decay of a color-singlet
$b\overline{b}(^1S_0)$ pair into $g\ g^*$, followed by $g^*\to c\
\overline{c}$\ .

We can calculate the process in the nonrelativistic limit, in which
the $b\overline{b}(^1S_0)$ pair can be taken as no relative momentum
within $\eta_b$, i.e., $p_b=p_{\overline b}\ =\ Q/2$, where Q is the
momentum of the $\eta_b$. In this situation, for the $b
\overline{b}$ pair to form $\eta_b$, when it is in a color-singlet
state, one can replace the product of the Dirac spinors for $b$ and
$\overline{b}$ in the initial state with the projector:
\bqa
u(p_b)\,\overline{v}(p_{\bar b})& \longrightarrow& {1\over 2
\sqrt{4\pi}} \,(\not\! Q+ M_{\eta_b})\,i\gamma_5 \times \left(
{1\over \sqrt{M_{\eta_b}}} R_{\eta_b}(0)\right) \otimes \left( {{\bf
1}_c\over \sqrt{N_c}}\right)\ , \label{Etab:projector} \eqa
where $N_c = 3$, and $1_c$ stands for the unit color matrix.
$M_{\eta_b}$ is the masses of $\eta_b$. At leading order in
non-relativistic expansion, it could be understood that $M_{\eta_b}\
\approx \ 2\ m_b$. The nonperturbative parameters, $R(0)_{\eta_b}$
are color-single radial wave functions at the origin for $\eta_b$,
which can be either reached from phenomenological potential models
or directly extracted from experiments. The relation between the
$R(0)_{\eta_b}$ and $\langle\mathcal{O}^{\eta_b}_1(^3S_1)\rangle$
reads $\langle\mathcal{O}^{\eta_b}_1(^3S_1)\rangle=(N_c/2\pi)
|R(0)_{\eta_b}|^2(1+O(\upsilon^4))$.

One can then get the partial decay width straightforwardly for the
process $b\bar{b}(^1S_0)\to c(p_1)+\bar{c}(p_2)+g(k)$. That is,
\begin{eqnarray}
d\Gamma[b\bar{b}(^1S_0)\to
c\bar{c}g]=\frac{1}{2M_{\eta_b}}\sum_{ccg} |M_{str}|^2\ d\Phi_3\ ,
\end{eqnarray}
where $M_{str}$ is the amplitude for the process and the $\Phi_3$
represents the three-body phase space, which shows
\begin{eqnarray}
d\Phi_3&=&\frac{1}{(2\pi)^9}\cdot\frac{d^3{p_1}}{2E_1}\cdot\frac{d^3{p_2}}{2E_2}\cdot
\frac{d^3{k}}{2k_0}\cdot(2\pi)^4\delta^4(Q-p_1-p_2-k)\ .
\end{eqnarray}
In the initial state rest frame, after integrating over the
variables which are independent of the amplitude , the phase space
integration can be further simplified as
\begin{eqnarray}
d\Phi_3^\prime=\frac{1}{(2\pi)^3}\frac{1}{4M_{\eta_b}}dE_1dS_{13}\ .
\end{eqnarray}
Here, $S_{13}=p_1\cdot k$\ .

In the numerical calculation, we take $m_b = 4.65\ \pm\ 0.15\ {\rm
GeV}$, $m_c = 1.50\ \pm\ 0.05\ {\rm GeV}$, and $\alpha_s(m_b) =
0.22$. The magnitude of the radial wave function at the origin
$R(0)$ of $\eta_b$ equals approximately to that of its spin triplet
partner $\Upsilon$, which can be determined from the experimental
data on the decay width of $\Gamma(\Upsilon\to
e^+e^-)=(1.340\pm0.018)\times10^{-6} \rm GeV$ \cite{PDG}.  That is:
$|R(0)_{\eta_b}|^2 = |R(0)_{\Upsilon}|^2 = 4.89\pm0.07$ ${\rm
GeV}^{3}$. With these inputs and by varying the strong coupling
scale from $m_b/2$ to $2 m_b$, for the aim of error estimation, we
have
\begin{eqnarray}
\Gamma[b\bar{b}(^1S_0)\to c\bar{c}g] = 190.8^{+190.0}_{-86.3}\ {\rm
KeV}\ .
\end{eqnarray}
Since at leading order the total decay width of $\eta_b$ is
\bqa \Gamma_{tot}(\eta_b)\approx\Gamma(\eta_b\rightarrow
gg)=\frac{8}{3}\frac{\alpha_s^2}{M_{\eta_b}^2}| R_{\eta_b}(0)|^2\
,\eqa
the branching ratio hence is readily obtained to be
\begin{eqnarray}
Br[b\bar{b}(^1S_0)\to c\bar{c}g]= 2.6^{+0.9}_{-0.6}\times 10^{-2}\ .
\end{eqnarray}

\begin{figure}
\begin{center}
\includegraphics[scale=0.9]{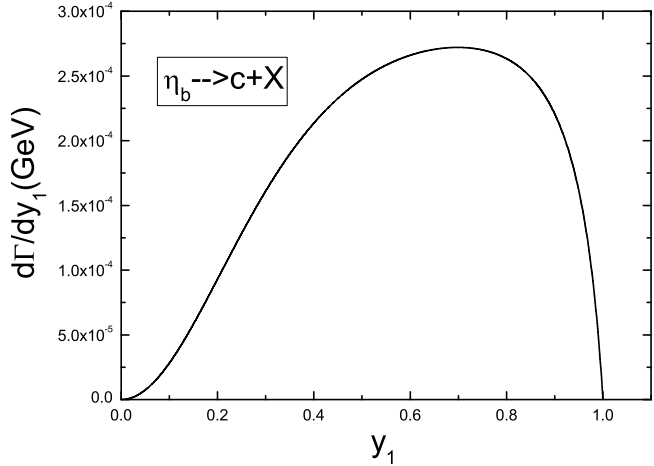}
\caption{The decay rate variation over momentum fraction $y_1$ in
the inclusive process $\eta_b \rightarrow c + X$, by taking the
central values of inputs.
 \label{Diag-y1}}
\end{center}
\end{figure}

For inclusive decay process, giving out the experiment observable
differential distribution will be useful. For this purpose, We
define two fractions $x_1=E_1/E_b$ and $r_c=m_c^2/m_b^2$, where
$E_1$ and $E_b$ stand for the energy of charm quark and bottom
quark, and in nonrelativitic approximation the $E_b = m_b =
M_{\eta_b}/2$. The region of variable $x_1$ is $\sqrt{r_c}< x_1 <1$.
In some cases, instead of $x_1$ it is convenient to use another
variable $y_1$, which is the momentum of the charm quark divided by
its kinematically allowed maximum value in $\eta_b$ decays
\cite{CPX,CPW}. The relation between these two variables reads
\begin{eqnarray}
x_1&=&\sqrt{(1-r_c)y_1^2+r_c}\ , \\
y_1&=&\sqrt{\frac{x_1^2-r_c}{1-r_c}}\ .
\end{eqnarray}
The range of $y_1$ is $0<y_1<1$. Figure \ref{Diag-y1} exhibits the
decay rate distribution over the momentum fraction $y_1$.

Because almost all the charm quarks may eventually hadronize into
charmed hadrons, like $D^0$, $D^{\pm}$, $D_s$, and $\Lambda_c$,
etc., we schematically show the $D^+$ meson differential
distribution in $\eta_b$ decays in the fragmentation approximation,
similar as done in Refs. \cite{CPX,CPW}. It is well-known that the
fragmentation function $D_{c\to h}(z)$ represents the probability of
a charm quark fragmenting into the charmed hadron $h$. Here, the $z$
is a Lorentz boost invariant variable, defined as  $z =
\frac{E_h+p_h}{E_1+p_1}$. In practice calculation, we will simply
neglect the difference between fragmenting charm quark mass and the
charmed hadron mass. The $z$ can be reexpressed as:
\begin{eqnarray}
z=\frac{z_h}{z_1}
\end{eqnarray}
with
\begin{eqnarray}
z_1=\frac{\sqrt{(1-r_c)y_1^2+r_c}+ y_1\sqrt{1-r_c}}{1+\sqrt{1-r_c}}\ ,\\
z_h=\frac{\sqrt{(1-r_c)y_h^2+r_c}+ y_h
\sqrt{1-r_c}}{1+\sqrt{1-r_c}}\ .
\end{eqnarray}
Then the momentum distribution of the charmed hadron can be
expressed as \cite{CPX,CPW}
\begin{eqnarray}
\frac{d\Gamma}{dy_h}&=&\frac{dz_h}{dy_h}\int^1_{z_h}
\frac{dz_1}{z_1}D(z_h/z_1)\frac{dy_1}{d z_1}\frac{d\Gamma}{dy_1}\ ,
\end{eqnarray}
According to the Kartvelishvili-Likhoded-Petrov (KLP) \cite{FGF}
fitting for fragmentation function
\begin{eqnarray}
D_{c\to h}(z)=N_h z^{\alpha_c}(1-z)\ . \label{frag}
\end{eqnarray}
Here, by using the optimal value of $\alpha_c = 4$ for $D^+$ fitted
by the Belle Collaboration \cite{Belle00}, one gets the
normalization coefficient $N_h = 8.04$ \cite{CPX}.

\begin{figure}
\begin{center}
\includegraphics[scale=0.9]{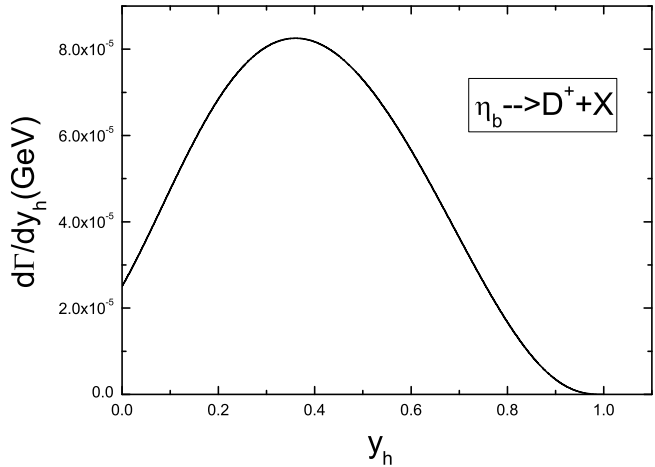}
\caption{The decay rate variation over momentum fraction $y_h$ in
the inclusive process $\eta_b \rightarrow D^+ + X$ with central
values of inputs.
 \label{Diag-dd}}
\end{center}
\end{figure}

We show in Figure \ref{Diag-dd} the $D^+$ meson differential
production distribution in $\eta_b$ decays in the fragmentation
approximation. For other charmed hadrons, the corresponding
distributions can be obtained similarly.

It should be mentioned that since the NRQCD factorization breaks
down in the $y_1 \rightarrow 1$ limit, the velocity and coupling
expansions are no more valid in the vicinity of the end point. A
proper treatment for this endpoint illness is to resum the large
logarithms of $\log(1 - y_1)$ \cite{Bauer:2000ew,Bauer:2001rh,
Fleming:2002rv,Fleming:2002sr,Fleming:2004rk,Fleming:2004hc} and
invoke the shape function \cite{Beneke:1997qw}. For these kinds of
content readers should refer to the related references in the
literature; and in this situation, the final state distribution
results at the endpoint in this work should not be taken seriously.
Roughly speaking, when $y_1 < 0.7$ the endpoint effects become weak
and our predictions turn to be robust \cite{GarciaiTormo:2005ch}.
Fortunately, for total decay widths, the endpoint influence is minor
and the predictions are quite reliable. For the charmed hadron
production, apart from the upper point problem, the lower limit also
poses a problem for the fragmentation approximation, although
(\ref{frag}) only comes from the phenomenological fitting.
Therefore, the lower endpoint prediction should not be taken
seriously as well. In Ref.\cite{CPX}, there are detailed discussions
about the validity of the fragmentation approximation.

\section{$\eta_b$ inclusive decay to $J/\psi + X$}\label{section3}

\begin{figure}
\begin{center}
\includegraphics[scale=0.9]{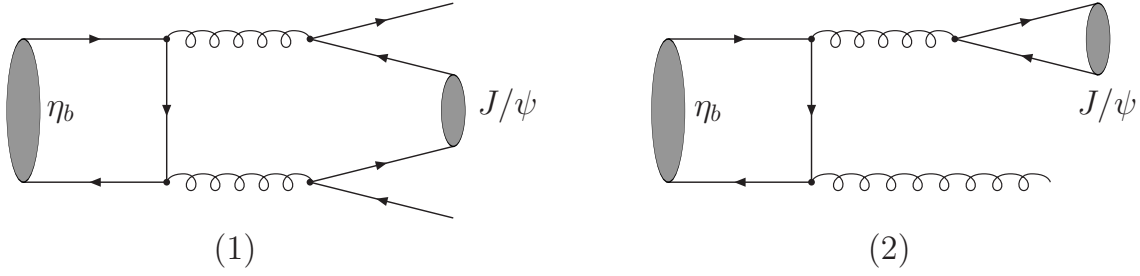}
\caption{Lowest-order diagrams that contribute to the inclusive
process: $\eta_b\rightarrow J/\psi\,X$.  The $J/\Psi$ in Diagrams
(1) is a color-single state , while the one in (2) is a color-octet
state. \label{Diag-fm}}
\end{center}
\end{figure}

In this section, we present the calculation of $J/\psi$ inclusive
production and its momentum distribution in the $\eta_b$ decays, as
shown in Figure \ref{Diag-fm}. As mentioned in above, at the leading
order in $\upsilon$ only the color-singlet $b\bar{b}(^1S_0)$
contribution in the initial state is the necessary ingredient to be
considered, since the higher Fock state contribution can not get big
enhancement from kinematic or dynamic reasons. However, for the
final state $J/\psi$ both the color-singlet and color-octet effects
should be included, because the latter one can get one $\alpha_s$
and kinematic compensation in comparison with the color-singlet
process. Since it is the $b\bar{b}$ pair annihilation in the initial
state and the creation of the $c \bar{c}$ pair, as well as the
emission of gluon or quarks, takes place at the hard scales set by
the heavy quark masses, it is legitimate to tackle this
semi-inclusive process in the pQCD framework.

The NRQCD formalism allows the systematic calculation of inclusive
cross sections for quarkonium decays in perturbation QCD to any
order in $\alpha_s$ and $v^2$ , where $v$ is the typical relative
velocity of the heavy quark inside the quarkonium. Using the NRQCD
velocity scaling rules \cite{NRQCD}, we know the color-singlet
process, the Figure \ref{Diag-fm}(1), is at order of $\alpha^4_s
v^3$, since the matrix element $\langle O^V_1(^3S_1) \rangle$ is of
order $v^3$. Whereas, the color-octet process, the Figure
\ref{Diag-fm}(2), is at order $\alpha^3_s v^7$. Here, $v$ denotes
the heavy quark relative velocity in Charmonium system. Potential
model calculation indicates that the average value of $\upsilon^2$
is about 0.3. And, the QCD coupling constant
$\alpha_s(m_b)\approx0.22$. In addition of the $\alpha_s$
compensation, the color-octet mechanism may also enhanced by the
single gluon propagator. Our following calculation really shows that
the color-octet process should be included in this calculation.

The decay width can be formulated as
\begin{eqnarray}
\Gamma[\eta_b\to J/\psi+X]=A_1\langle O^{J/\psi}_1(^3S_1)\rangle
+A_2\langle O^{J/\psi}_8(^3S_1)\rangle \ ,
\end{eqnarray}
where, the $A_1$ and $A_2$ are perturbative calculable
short-distance coefficients. We first calculate the color-single
coefficient. It is customary to start with the parton process; here
it is $b(p_b)\,\bar{b}(p_{\bar b}) \to c(p_c) \,\bar{c}(p_{\bar{c}})
+ c(k_1)+\overline{c}(k_2)$. Then project the matrix element onto
corresponding color-singlet configurations. For the $b\overline{b}$
in initial state, it is the same as in the last section. For the
$J/\psi$ production, the color-singlet projector is
\bqa
v(p_{\bar c})\,\overline{u}(p_c)& \longrightarrow& {1\over 2
\sqrt{4\pi}} \not\! \epsilon^*_{J/\psi}\,(\not\! P+M_{J/\psi})\,
\times \left( {1\over \sqrt{M_{J/\psi}}} R_{J/\psi}(0)\right)
\otimes \left( {{\bf 1}_c\over \sqrt{N_c}}\right)\ ,
\label{JPsi:projector} \eqa
where $\epsilon^{\mu}_{J/\psi}$ is the $J/\psi$ polarization vector
satisfying $\epsilon_{J/\psi}(\lambda)\cdot
\epsilon_{J/\psi}^*(\lambda^\prime)
=-\delta^{\lambda\lambda^\prime}$ and $P\cdot \epsilon_{J/\psi}=0$.
$R_{J/\psi}(0)$ is the radial wave function at the origin, which is
also valued through $J/\psi$ leptonic decay width. Combining all
these together, one can easily get the $\eta_b$ to $J/\psi + c +
\bar{c}$ decay amplitude for the color-single case, i.e.
\bqa M_{str}^1& =& C_1g_s^4\frac{R_{\eta_b}(0)R_{J/\psi}(0)}
{4\pi\sqrt{M_{J/\psi}M_{\eta_b}}} \nn\nonumber\\&\times& Tr[(\not\!
Q+ M_{\eta_b})\gamma_5\gamma_\mu((k_2-k_1)
\cdot\gamma+M_{\eta_b})\gamma_\nu]\nn
\nonumber\\&\times&\frac{1}{(k_2-k_1)^2-M_{\eta_b}^2}\times\frac{1}{(k_1+P/2)^2}
\times\frac{1}{(k_2+P/2)^2}\nn \nonumber\\
&\times&\overline{u}(k_1)\gamma^\mu\not\! \epsilon_{J/\psi}(\not\!
P+M_{J/\psi})\gamma^\nu v(k_2)\ ,\eqa
where $C_1$ is the corresponding color factor .  $k_1$ and $k_2$ are
the momenta carried by the external charm quark and anti-quark,
respectively.

Next, we present the calculation for color-octet process. At the
parton level it is $b(p_b)\,\bar{b}(p_{\bar b}) \to c(p_c)
\,\bar{c}(p_{\bar{c}}) + g(k)$ process followed by projecting the
$c$ $\bar{c}$ spinors onto the color-octet configuration,
$^3S_1^{[8]}$, while keeping on configuring the initial $b$
$\bar{b}$ in color-singlet. The color-octet projector is
\bqa
v(p_{\bar c})\,\overline{u}(p_c)& \longrightarrow& {1\over 2
\sqrt{4\pi}} \not\! \epsilon^*_{J/\psi}\,(\not\! P+M_{J/\psi})\,
\times \left( {1\over \sqrt{M_{J/\psi}}} R^8_{J/\psi}(0)\right)
\otimes\sqrt{2}\ T^a_{ij}\ . \label{JPsi:projector} \eqa
Here, $T^a_{ij}$ denotes the SU(3) generator. And, we introduce
another phenomenological parameter $R^8_{J/\psi}(0)$, which stands
for the color-octet nonperturbative effect. The relation between
$R^8_{J/\psi}(0)$ and the NRQCD matrix element $\langle
\mathcal{O}^{J/\psi}_8(^3S_1) \rangle$ is defined as:
\begin{eqnarray}
\langle \mathcal{O}^{J/\psi}_8(^3S_1) \rangle=\frac{3N_c}{2\pi}|
R^8_{J/\psi}(0)|^2\ .
\end{eqnarray}
From the fitted value of $\langle \mathcal{O}^{J/\psi}_8(^3S_1)
\rangle \approx 1.5\rm\times10^{-2}GeV^3 $ \cite{8CM}, we have $|
R^8_{J/\psi}(0)|=0.102\ \rm GeV^{3/2}$. Then the decay amplitude for
color-octet case is
\bqa M_{str}^{8(a)}(\lambda_1,\lambda_2) & = &
C_8g_s^3\frac{R_{\eta_b}
(0)R^8_{J/\psi}(0)}{4\pi\sqrt{M_{J/\psi}M_{\eta_b}}}
\nn\nonumber\\&\times& Tr[(\not\! Q+
M_{\eta_b})\gamma_5\not\!\epsilon_g^{\,a}(\lambda_2)((Q/2-k)
\cdot\gamma+M_{\eta_b}/2)\gamma^\nu]\nn
\nonumber\\&\times&\frac{1}{M_{J/\psi}^2}
\times\frac{1}{M_{\eta_b}^2-M_{J/\psi}^2} \times Tr[\not\!
\epsilon_{J/\psi}(\lambda_1)(\not\! P+M_{J/\psi})\gamma_\nu]\ ,\eqa
where $C_8$ is the color factor, $k$ is the momentum carried by the
external gluon, and $\epsilon^{\mu}_g$ is the gluon polarization
satisfying $k\cdot \epsilon_g=0 $.

With the matrix elements $M_{str}^1$ and $M_{str}^8$ we can
immediately get the $\eta_b \rightarrow J/\psi + X$ decay width. The
analytical result for it is a bit lengthy, and will not be presented
here. For our numerical estimation, the non-relativistic limit is
also enforced for the charmonium. That is we take the $M_{J/\psi}
\approx 2 m_c$ approximation. From $\Gamma(J/\psi \rightarrow
e^+e^-)=(5.55\pm0.14)\times10^{-6} \rm GeV$ \cite{PDG}, we get
$|R_{J/\psi}(0)|^2=0.527 \pm 0.013\ \rm GeV^3$. Then, the decay
width for the concerned process reads,
\bqa \Gamma(\eta_b\rightarrow
J/\psi_{color-singlet}+X)&=&0.13^{+0.26}_{-0.08}\ \rm\ KeV\ , \\
\Gamma(\eta_b\rightarrow
J/\psi_{color-octet}+X)&\cong&2.16\ \rm\ KeV\ , \\
\Gamma_{total}(\eta_b\rightarrow J/\psi+X)&\cong& 2.29\ \rm\ KeV\ .
\eqa
That means that the $\eta_b\rightarrow J/\psi+X$ process has a
branching ratio of $3.23\times10^{-4}$ or so in the $\eta_b$ decays.
Here, the uncertainty estimate of the color-singlet process is
performed in the same way as in the preceding section. Whereas,
considering of the large uncertainties remaining in the color-octet
matrix element fitting, we carry the numerical calculation for
color-octet process by only taking the central values of the inputs.

Like in section II, to give out the differential decay width we
define three new variables, the $x_2$, $r_{J/\psi}$, and $y_2$ , as
\bqa x_2&=&E_{J/\psi}/E_b\ , \\
r_{J/\psi}&=&M_{J/\psi}^2/m_b^2\ , \\
y_2&=&\sqrt{\frac{x_2^2-r_{J/\psi}}{1-r_{J/\psi}}}\ .
 \eqa
Then we can express the partial decay width of $\eta_b\rightarrow
J/\psi_{color-singlet} + X$ as
\begin{eqnarray}
 d\Gamma[\eta_b\rightarrow
J/\psi_{color-single}+X]=\frac{1}{2M_{\eta_b}} \sum_{J/\psi
 c\overline{c}}| M_{str}^1|^2
\ d\Phi_3
\end{eqnarray}
\begin{figure}
\begin{center}
\includegraphics[scale=0.9]{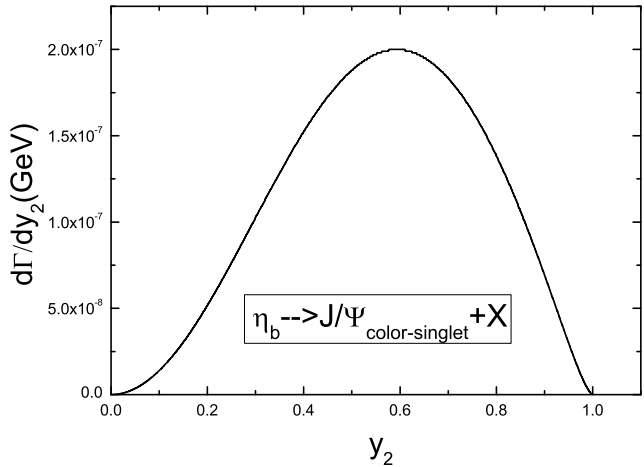}
\caption{The $J/\psi$ momentum distribution in the inclusive process
$\eta_b\rightarrow J/\psi_{color-singlet}+X$.
 \label{Diag-jp}}
\end{center}
\end{figure}

In analogy to what is performed in the last section, we get the
momentum distribution $d\Gamma(\eta_b\rightarrow
J/\psi_{color-single}+X)/dy_2$, as showen in Figure \ref{Diag-jp}.
For the the process $\eta_b\rightarrow J/\psi_{color-octet}+ X$,
\bqa d\Gamma[\eta_b\rightarrow
J/\psi_{color-octet}+X]=\frac{1}{2M_{\eta_b}}\sum_{\lambda_1,\lambda_2}|
M_{str}^8(\lambda_1,\lambda_2)|^2d\Phi_2\ . \eqa
Since this is a two-body decay process, the $J/\psi$ momentum
distribution $d\Gamma(\eta_b\rightarrow
J/\psi_{color-octet}+X)/dy_2$ is only a delta function peaked at
$y_2=\sqrt{\frac{(M_{\eta_b}^2-M_{J/\psi}^2)^2}
{4M_{\eta_b}^2(M_{\eta_b}^2-4M_{J/\psi}^2)}}=0.6$. Again, for the
$\eta_b$ to $J/\psi$ inclusive decay distribution, one should pay
attention to the endpoint problem \cite{LL}. In particular for the
color-octet contribution, the $\eta_b$ two-body decay at leading
order resulting in a delta function distribution, which is smeared
out by the non-perturbative effects and resulting in a shape
function \cite{FLM1}.

The numerical result shows that the branching ratio for color-octet
process is larger than the one for color-singlet process by about an
order, which offers an opportunity to check the existence of
color-octet mechanism experimentally. Considering of the
uncertainties existing in the magnitude of color-octet matrix
element, the numerical difference between these two processes might
shrink down, nevertheless, they give a distinctively different
momentum distribution, which may also help experimenters to
distinguish them in the future experiment.

\section{summary}\label{section4}

We have studied in the framework of NRQCD the inclusive charm
production in the decay of the pseudoscalar bottomnium state
$\eta_b$. We find that it gives a quite large branching fraction.
Since the produced charm quarks will dominantly evolve into charmed
hadrons, by employing the KLP fragmentation function, we give out
the momentum distribution of $D^+$, as an example.

We have also calculated the decay width and the momentum
distribution of the inclusive $J/\psi$ production in the $\eta_b$
decay. We find that in this case the color-octet process should be
taken into consideration. However, this two different $J/\psi$
production schemes have obviously different momentum distributions.
This is a distinct character of this process, which will be helpful
for future experiment to investigate the $\eta_b$ and to the study
$J/\psi$ production.

In all, to investigate the elusive $\eta_b$ is still an interesting
task for both theory and experiment. Our explicit calculation shows
that $\eta_b$ inclusive decays to charm pair(in experiment the
charmed hadron pair) and $J/\psi$ have quite large branching
fractions. These processes can be helpful for people to hunt for the
$\eta_b$ at the Fermilab Tevatron, or LHC, where copious $\eta_b$
are expected.
\vspace{0.8cm}
\par
\par

{\bf Acknowledgments} \vspace{.2cm}

This work is supported in part by the National Natural Science
Foundation of China, by the Scientific Research Fund of GUCAS (NO.
055101BM03), and by fund provided by the U.S. Department of
Energy(D.O.E) under cooperative research agreement DEFG02-05ER41360
under Particle.

\end{document}